

\input lanlmac

\def\/{\over}
\def\'{\prime}
\def\t{\widetilde}
\def\to{\rightarrow}
\def\d{\partial}
\def\pder#1{{\partial\over\partial{#1}}}

\def\hf{{1 \over 2}}

\def\sitarel#1#2{\mathrel{\mathop{\kern0pt #1}\limits_{#2}}}
\def\inbar{\,\vrule height1.5ex width.4pt depth0pt}
\def\IC{\relax\hbox{$\inbar\kern-.3em{\rm C}$}}
\def\IR{\relax{\rm I\kern-.18em R}}
\def\IP{\relax{\rm I\kern-.18em P}}

\def\R{{\bf R}}
\def\S{{\bf S}}
\def\T{{\bf T}}

\def\1{{1\hskip -3pt {\rm l}}}
\def\one{{1\hskip -3pt {\rm l}}}
\def\ie{{\it i.e.},$\ $}
\def\M{M_{\rm p}}
\def\a{\alpha'}
\def\l{l_{\rm s}}
\def\g{g_{\rm str}}
\def\tg{\tilde{g}_{\rm str}}
\def\frac#1#2{{{#1}\over{#2}}}
\def\ae{{\alpha'}_{\rm eff}}
\def\Me{M_{\rm eff}}

\def\G{G_{\rm o}}
\def\tG{\widetilde{G}_{\rm o}}
\def\Gs#1{G_{\rm {#1}}}
\def\tGs#1{\widetilde{G}_{\rm {#1}}}
\def\gs#1{\g^{(#1)}}
\def\tgs#1{\tg^{(#1)}}


\def\np#1#2#3{{ Nucl. Phys.} {\bf B#1}, #2 (#3)}

\def\npps#1#2#3{{ Nucl. Phys. Proc. Suppl.} {#1BC}, #2 (#3)}
\def\pln#1#2#3{{Phys. Lett. } {\bf B#1}, #2 (#3)}

\def\pr#1#2#3{{ Phys. Rev.} {\bf D#1}, #2 (#3)}
\def\prl#1#2#3{{ Phys. Rev. Lett.} {\bf #1}, #2 (#3) }

\def\jhep#1#2#3{{JHEP} {\bf #1}, #2 (#3)}

\def\hpt#1{{\tt hep-th/#1}}

\lref\GoMaMiSt{
R. Gopakumar, J. Maldacena, S. Minwalla, and A. Strominger, 
``$S$-duality and Noncommutative Gauge Theory,'' 
\hpt{0005048}. 
}
\lref\GoMiSeSt{
R. Gopakumar, S. Minwalla, N. Seiberg, and A. Strominger, 
``OM Theory in Diverse Dimensions,'' 
\hpt{0006062}. 
}
\lref\SW{
N. Seiberg and E. Witten, 
``String Theory and Noncommutative Geometry,'' 
\jhep{9909}{032}{1999}, 
\hpt{9908142}. 
}
\lref\Aspin{
P. S. Aspinwall, 
``Some Relationships between Dualities in String Theory,'' 
\npps{46}{30-38}{1996}, \hpt{9508154}.
}
\lref\John{
J. H. Schwarz, 
``The Power of M Theory,'' 
\pln{367}{97-103}{1996}, \hpt{9510086}.
}
\lref\BBSS{
E. Bergshoeff, D. S. Berman, J. P. van der Schaar,  and P. Sundell,
``Critical Fields on the M5-brane and Noncommutative Open Strings,''
\hpt{0006112}.
} 
\lref\SSTa{
N.~Seiberg, L.~Susskind, and N.~Toumbas,
``Strings in Background Electric Field, 
Space/Time Noncommutativity  and a New Noncritical String Theory,'' 
\jhep{0006}{021}{2000}, 
\hpt{0005040}. 
}
\lref\SSTb{
N.~Seiberg, L.~Susskind, and N.~Toumbas,
``Space/Time Non-Commutativity and Causality,''
\hpt{0005015}.
}
\lref\BR{
J.~L.~Barbon and E.~Rabinovici, 
``Stringy Fuzziness as the Custodian of Time-Space Noncommutativity,''
\hpt{0005073}.
}
\lref\GRS{
O.~J.~Ganor, G.~Rajesh, and S.~Sethi,
``Duality and Non-commutative Gauge Theory,''
\hpt{0005046}.
}
\lref\CDGR{
S.~Chakravarty, K.~Dasgupta, O.~J.~Ganor, and G.~Rajesh,
``Pinned Branes and New Non Lorentz Invariant Theories,''
\hpt{0002175}.
}
\lref\BBSSa{
E.~Bergshoeff, D.~S.~Berman, J.~P.~van der Schaar, and P.~Sundell,
``A Noncommutative M-theory Five-brane,''
\hpt{0005026}.
}
\lref\KS{
S.~Kawamoto and N.~Sasakura,
``Open Membranes in a Constant C-field Background 
and Noncommutative Boundary Strings,''
\hpt{0005123}.
}
\lref\HSWa{
P.~S.~Howe, E.~Sezgin, and P.~C.~West,
``Covariant Field Equations of the M-theory Five-brane,''
\pln{399}{49-59}{1997}, 
\hpt{9702008}.
}
\lref\HSWb{
P.~S.~Howe, E.~Sezgin, and P.~C.~West,
``The Six-dimensional Self-dual Tensor,''
\pln{400}{255-259}{1997}, 
\hpt{9702111}.
}
\lref\CW{
G.~Chen and Y.~Wu,
``Comments on Noncommutative Open String Theory: V-duality and Holography,''
\hpt{0006013}.
}
\lref\Harm{
T.~Harmark,
``Supergravity and Space-time Non-commutative Open String Theory,''
\hpt{0006023}.
}
\lref\KM{
I.~R.~Klebanov and J.~Maldacena,
``1+1 Dimensional NCOS and its U(N) Gauge Theory Dual,''
\hpt{0006085}.
}
\lref\GKP{
S. Gukov, I. R. Klebanov, and A. M. Polyakov, 
``Dynamics of $(N, 1)$ Strings,''
\pln{423}{64-70}{1998}, 
\hpt{9711112}.
}
\lref\Verlinde{
H. Verlinde, 
``A Matrix String Interpretation of the Large $N$ Loop Equation,'' 
\hpt{9705029}
} 
\lref\RS{
J. G. Russo and M. M. Sheikh-Jabbari, 
``On Noncommutative Open String Theories,''
\hpt{0006202}.
}
\lref\LRS{
J. X. Lu, S. Roy, and H. Singh, 
``((F, D1), D3) Bound State, S-duality 
and Noncommutative Open String/Yang-Mills Theory,''
\hpt{0006193}.
}
\lref\Berman{
D. Berman, 
``M5 on a Torus and the Three-brane,'' 
\np{533}{317-332}{1998}, 
\hpt{9804115}. 
}
\lref\GM{
J. Gomis and T. Mehen, 
``Space-time Noncommutative Field Theories and Unitarity,''
\hpt{0005129}. 
}
\lref\Pcski{
J. Polchinski,
``Dirichlet Branes and Ramond-Ramond Charges'',
\prl{75}{4724-4727}{95}, \hpt{9510017}.
}
\lref\Dbrane{
J. Polchinski, S. Chaudhuri, and C. V. Johnson, 
``Notes on D-Branes'',
\hpt{9602052}\semi
J. Polchinski, 
``TASI Lectures on D-Branes'',
\hpt{9611050}.
}
\lref\W{
E. Witten,
``Bound States Of Strings And $p$-Branes'',
\np{460}{335-350}{96}, \hpt{9510135}.
}
\lref\BFSS{
T. Banks, W. Fischler, S. H. Shenker, and L. Susskind,
``M Theory As A Matrix Model: A Conjecture'',
\pr{55}{5112-5128}{97}, \hpt{9610043}.
}
\lref\DH{
M.R.~Douglas and C.~Hull,
``D-Branes and the Noncommutative Torus,''
\jhep{02}{008}{98}, \hpt{9711165}\semi
Y.-K.E. Cheung and M. Krogh,
``Noncommutative Geometry from 0-branes in a Background B Field,''
\np{528}{185}{98}, \hpt{9803031}\semi
T. Kawano and K. Okuyama,
``Matrix Theory on Noncommutative Torus,''
\pln{433}{29}{98}, \hpt{9803044}.
}


\Title{                                \vbox{\hbox{UT-899}
                                             \hbox{\tt hep-th/0006225}} }
{\vbox{\centerline{
                      $S$-Duality from OM-Theory
}}}

\vskip .2in

\centerline{
                  Teruhiko Kawano and Seiji Terashima
}

\vskip .2in 


\centerline{\sl
               Department of Physics, University of Tokyo
}
\centerline{\sl
                     Hongo, Tokyo 113-0033, Japan
}
\centerline{\tt
                    kawano@hep-th.phys.s.u-tokyo.ac.jp
}
\vskip -0.05in
\centerline{\tt
                    seiji@hep-th.phys.s.u-tokyo.ac.jp
}

\vskip 3cm

We study the compactification of OM-theory on tori and show 
a simple heuristic derivation of the $S$-duals of noncommutative open 
string theory in diverse dimensions from the OM-theoretical point of view.  
In particular, we identify the $S$-duality between 
noncommutative open string theory and noncommutative Yang-Mills theory 
in $(3+1)$ dimensions as the exchange of 
two circles of a torus on which OM-theory is compactified. 
Also, we briefly discuss $T$-duality between noncommutative open string theories. 

\Date{June, 2000}


\newsec{Introduction}

Recently, in the papers \refs{\SSTa,\GoMaMiSt}, the authors found a new open 
string theory in which there were no closed strings. 
Let us consider a D-brane in a flat 
Minkowski space with a background $B$-field. We can see that open strings ending 
on the D-brane couple to the $B$-field, but that closed strings do not. 
When the temporal components of the 
background $B$-field are turned on, since strings spatially extend in one 
dimension, the $B$-field in the worldsheet action can contribute to 
the effective tension of open strings and can even make the tension 
vanish, if we tune the value of the temporal components appropriately. 
In the limit that the string tension goes to infinity (\ie 
$\a\rightarrow0$), we can keep the effective tension of open strings fixed, 
if the value of the $B$-field is arranged to be critical. 
Therefore, gravity decouples from the open string sector in the limit, where 
we still have all the exciting modes of the open strings. This open string theory 
is called noncommutative open string (NCOS) theory.  
Related topics have also been discussed in 
\refs{\SSTb\GRS\BR\CW\Harm\KM\GKP-\GM}.

Let us explain the NCOS limit in some detail. 
Suppose that a D$p$-brane is in a $p$-dimensional flat space with 
a background $B$-field and a metric 
\eqn\metric{
g_{\mu\nu}=\eta_{\mu\nu}\quad(\mu,\nu=0,1),
\qquad
g_{ij}=f^2\delta_{ij}\quad(i,j=2,\ldots,p),
}
and $f\to0$ in the NCOS limit. 
By using a gauge symmetry, under which the gauge field $A$ 
on the D-brane and the $B$-field $B$ transform as $A\rightarrow{A-\Lambda}$, 
$B\rightarrow{B+d\Lambda}$, we make the background $B$-field vanish and, 
instead, obtain the corresponding background field strength $F=dA$. 
In order to take 
the field strength to the critical value in the limit, we set 
\eqn\Ecr{
2\pi\a F_{01}=1-\hf f^2. 
}
Translating these parameters into the open string moduli in \SW, we find that 
the noncommutative parameter $\Theta^{01}$ becomes $2\pi\ae$, where $\ae$ is 
related to $f$ as $f^2=(\a/\ae)$. 
Since the open string metric $G_{AB}$ ($A,B=0,\ldots,p$) equals $f^2\eta_{AB}$, 
the open string coupling constant $\G^2$, defined as 
$\g\sqrt{-\det{G}}/\sqrt{-\det(g+2\pi\a F)}$, turns out to be ${\g}f$. 
The NCOS limit is defined as $\a\rightarrow0$ ($f\rightarrow0$), 
$\g\rightarrow\infty$ with $\ae$ and $\G^2$ fixed. 
In the limit, the effective string tension becomes $1/(2\pi\ae)$, which
gives the scale of the exciting modes of open strings.

In the paper \GoMiSeSt, it was proposed that OM-theory describes 
the strong coupling limit of $(4+1)$-dimensional NCOS theory (NCOS$_{4+1}$). 
The relation of OM-theory to NCOS$_{4+1}$ is similar to that of M-theory 
to type IIA superstring theory. 
The effective tension of open membranes within M5-branes 
with nonvanishing temporal components of the background three-form 
tensor field can be different from the tension $\M^3/(2\pi)^2$ of membranes 
where $\M$ is the eleven-dimensional gravitational scale. 
In the limit $\M\rightarrow\infty$, gravity would decouple from the open 
membranes, if the effective tension is kept finite. 
In this way, OM-theory is defined as the limit of this system.

In order to elucidate OM-theory more precisely, we begin with 
a single M5-brane in a flat Minkowski spacetime where we have the metric 
\metric\ with $p=5$ and a background worldvolume three-form field strength $H$ 
\eqn\He{
H_{012}=\M^3 \, \tanh\beta. 
}
Here $f^2=2\exp(-\beta)/\cosh\beta=2\Me^3/\M^3$, where $\Me^3/(2\pi)^2$ is 
the effective tension of open membranes in the OM limit. The self-duality 
of the field $H$ means, as discussed in \HSWb, that 
\eqn\Hm{
H_{345}=\Me^3f\sinh\beta. 
}
Now we can define 
OM-theory by the limit $\M\rightarrow\infty$ with $\Me$ fixed. 
In this limit, open membranes (M2-branes) stretched on the $x^1$-$x^2$ plane 
or on the $x^i$-$x^j$ plane ($i, j=3,4,5$) have a finite effective tension. 

NCOS$_{4+1}$ can be obtained by the compactification of OM-theory on $S^1$ 
where the coordinate $x^2$ is periodically identified and the radius is $R_2$. 
As we will see in the next section, 
by exploiting the relation between an M5-brane wrapped on $S^1$ in M-theory 
and a D4-brane in type IIA superstring theory, the open string moduli is given 
in terms of the parameters $R_2$, $\Me$ as $1/\ae=2R_2\Me^3$ and 
$\G^2=\sqrt{2}(R_2\Me)^{3\/2}$. 
Thus, the strong coupling limit $\G\rightarrow\infty$ of NCOS$_{4+1}$ can be seen 
to be the decompactification (\ie $R_2\to\infty$) of the OM-theory on $S^1$, 
and NCOS$_{4+1}$ becomes (5+1)-dimensional OM-theory in the limit. 

In \GoMaMiSt, the strong coupling limit of (3+1)-dimensional NCOS theory has 
also been discussed. Since NCOS$_{3+1}$ is given by the NCOS limit of 
a D3-brane in type IIB superstring theory, by $S$-duality of type IIB 
superstring theory we find that the strong coupling behavior of NCOS$_{3+1}$ 
is described by (3+1)-dimensional spatially noncommutative Yang-Mills theory. 

Aspinwall \Aspin\ and Schwarz \John\ interpreted $S$-duality of the 
nine-dimensional type IIB superstring theory as the modular transformation 
of a torus on which M-theory is compactified; see also \Berman.
This is one of many examples where M-theory plays an important role to unify 
superstring theories and eleven-dimensional supergravity. 
In this paper, by applying their idea to OM-theory compactified on a 
two-dimensional torus, we give a simple derivation of $S$-duality between 
spatially noncommutative Yang-Mills theory and NCOS theory in (3+1) dimensions. 
To this end, we show that the same OM limit in this case can be identified as 
both of the NCOS limit and the NCYM limit. Then, as Aspinwall and Schwarz gave 
the closed string coupling in terms of the moduli of the torus, we obtain the 
relation between the {\it open} string couplings in NCYM and NCOS by making use 
of the underlying torus. 

Furthermore, we also discuss from the OM-theoretical point of view, 
the compactification of OM-theory on higher dimensional tori to give 
a heuristic derivation of the $S$-duals of NCOS theories in diverse dimensions. 
The authors of the paper \GoMiSeSt\ have already identified the $S$-dualized 
theories. NCOS theory in (1+1) dimensions has been discussed in more detail 
in \refs{\KM}. 
We use an M2-brane to discuss the $S$-duality of this theory in section 3. 
We also discuss $T$-duality between NCOS theories in section 4.

\newsec{$S$-duality from OM-theory}

\subsec{Four-dimensional $S$-duality from OM-theory}

In this subsection, we consider OM-theory compactified on a rectangular 
torus, where the coordinates $x^2$ and $x^3$ are periodically identified 
as $x^2\sim x^2+2\pi R_2$ and $x^3\sim x^3+2\pi L_3$, and 
their radii are $R_2$ and $R_3=fL_3$, respectively.
Note that radius $R_3$ is much smaller than $R_2$. 
The compactification of M-theory on a circle of radius $R$ gives 
type IIA superstring theory with 
the string coupling constant $\g=(R \M)^{3\/2}$ and the string length $\l$ 
given by $\l^2=\a=(1/R \M^3)$. M5-branes wrapped on this circle become 
D4-branes in type IIA theory. 
Since we have two circles in the $x^2$ and $x^3$ directions in the torus, 
two IIA theories can be obtained by deciding 
which circle to identify as the above circle of radius $R$. 
$T$-duality then maps the two IIA theories into two corresponding IIB theories 
in nine dimensions. 
These IIB theories are found to be $S$-dual to each other, 
as Aspinwall and Schwarz have shown in \refs{\Aspin,\John,\Berman}. 

First, identifying $R_2$ as $R$, we regard the wrapped M5-brane as a D4-brane 
in IIA theory with the string coupling constant $\gs{2}=(R_2 \M)^{3\/2}$ and 
the string tension 
\eqn\Tf{
\frac{1}{{\alpha'}_2}= R_2 \M^3=2f^{-2}R_2\Me^3
}
on the circle of the radius $R_3=fL_3$.
The field strength of the gauge field on the D4-brane worldvolume
has a nearly critical electric component 
\eqn\Fe{
F_{01}=\frac{1}{2 \pi} R_2 H_{012} =
\frac{1}{2 \pi {\alpha'}_2} {\tanh \beta}
\sim \frac{1}{2 \pi {\alpha'}_2} 
\left( 1 -\hf f^2 \right).
}
If we take the limit $R_3\to\infty$,
this theory turns into $(4+1)$-dimensional NCOS theory with 
the effective string tension 
\eqn\aee{
\frac{1}{\ae^{(2)}}=2R_2 \Me^3 
}
and the open string coupling constant 
$\G^2 =\sqrt{2} (R_2 \Me)^{\frac{3}{2}}$, 
as shown in \GoMiSeSt.

Under the $T$-duality on the circle of the finite radius $R_3$, 
the D4-brane is mapped to a D3-brane in type IIB theory compactified on 
a circle of radius 
\eqn\tRe{
\t R_3=\frac{\alpha'_2}{ R_3} =
\frac{f}{2\Me^3R_2 L_3 },
}
where the string coupling constant is 
\eqn\tge{
\tgs{2}=\gs{2} \left(\frac{\t R_3}{R_3 }\right)^\hf
= \frac{R_2}{f\,L_3}.
}
The open string metric $G_{AB}$ $(A,B=0,1,4,5)$ and the noncommutative parameter 
$\Theta^{AB}$ on the D3-brane are given as in \SW\ by $G_{AB}=f^2\eta_{AB}$ and 
$\Theta^{01}=2\pi\ae^{(2)}$. The open string coupling constant turns out to be
\eqn\oge{
\tGs{E}^2 =\frac{R_2}{L_3}.
}

Thus, we can see that 
the NCOS limit in NCOS$_{3+1}$ agrees with the OM limit with $R_2$ and $L_3$ 
fixed. Note that, since the mass of D-strings wound once on the circle is 
\eqn\tTone{
M_{\rm D1} \sim \frac{1}{ \tgs{2}} \frac{\t R_3 }{\alpha'_2}=
\frac{1}{R_2} 
}
and remains finite in the OM limit (\ie $\M\to\infty$ with $\Me$ fixed), 
the D-strings are not decoupled from the worldvolume theory,
although the tension of D-strings is very large: 
$ T_{\rm D1} \sim {1/(f\tGs{E}^2\ae^{(2)}) } \gg {1/\ae^{(2)}}$. 
Thus, to discard those additional degrees of freedom and 
obtain (3+1)-dimensional NCOS theory, after taking the OM limit 
we should take the limit $R_2\to0$. In order to keep the effective 
string tension $1/\ae^{(2)}$ and the open string coupling constant 
$\tGs{E}$ fixed in the limit, we have to take the limit $\Me^3\to\infty$ and 
$L_3\to0$ at the same time.

Next, let us think of the M5-brane on the torus
as a D4-brane in another type IIA theory 
with the string coupling constant $\gs{3}=(R_3 \M)^{\frac{3}{2}}$ and 
\eqn\aem{
{{\alpha'}_3}= {f\/2L_3\Me^3}
}
on the circle of the radius $R_2$. 
In this picture, the field strength on the worldvolume
has a nonzero magnetic component 
\eqn\Fm{
2\pi F_{45}=R_3 H_{345} 
\sim 2L_3\Me^3.
}

By $T$-duality on the circle in the $x^2$ direction,
we obtain a D$3$-brane in type IIB theory compactified on a circle 
of radius 
\eqn\tRm{
\t{R}_2=\frac{\alpha'_3 }{ R_2} =
\frac{f}{2R_2L_3\Me^3}, 
}
with the string coupling constant
\eqn\tgm{
\tgs{3}={\gs{3}} \left(\frac{R'_2}{R_2 } \right)^\hf
= f\frac{L_3}{R_2}
=\frac{1}{\tgs{2}}.
}
The open string coupling constant becomes 
\eqn\ogm{
\tGs{M}^2= \tgs{3}
\sqrt{\frac{{\det}{G}}{{\det}(g+ 2\pi \alpha'_2 F)}}
=\frac{L_3}{R_2}.
}
The open string metric and the noncommutative parameter are given by 
$G_{MN}=\eta_{MN}$ $(M, N=0,1,4,5)$ and 
\eqn\NCtheta{
\Theta^{45}={\pi\/L_3\Me^3}.
} 
From \metric, \aem, and \Fm, 
we can verify that the OM limit with 
$R_2$ and $L_3$ fixed is the same limit as the NCYM limit found in 
\refs{\SW,\GoMiSeSt}. Note that, in this limit, the string coupling constant 
$\tgs{3}$ goes to zero. Therefore, the perturbative IIB string theory gives 
a good description in the limit. 
However, as the wound D-strings in the previous case are $S$-dual to 
open strings wound on the circle in the $x^2$ direction, 
the mass of the open strings is $M_{\rm F1} \sim R'_2/\alpha'_2
={1/R_2}$, and we obtain extra degrees of freedom from the wound open 
strings in the limit, besides those in (3+1)-dimensional noncommutative 
Yang-Mills theory (NCYM). 
Therefore, to decouple these degrees of freedom, 
we need to take $R_2$ to zero. In this limit, since we want to keep 
$\tGs{M}$ and $\Theta^{45}$ fixed, we have to take the 
limit $L_3\to0$ and $\Me\to\infty$. 
Thus, the theory on the D3-brane becomes (3+1)-dimensional 
noncommutative Yang-Mills theory (NCYM) in the above limit. 

This limit is in agreement with that in the previous case used 
to obtain NCOS$_{3+1}$. Comparing the open string coupling constant 
$\tGs{E}$ in \oge\ with $\tGs{M}$ in \ogm\ shows that
\eqn\Sduality{
\tGs{E}={1\/\tGs{M}}
}
and that the exchange 
$R_2\leftrightarrow L_3$ corresponds to $S$-duality between NCOS 
theory and NCYM theory in (3+1) dimensions. 
We also notice that the fluxes of the electric field $F_{01}$ and 
the magnetic field $F_{45}$ come from the same background self-dual $H_{012}$ 
and are indeed unified in OM-theory, 
although they seem very different in lower dimensions.

In the NCYM$_{3+1}$, as a consistency check, let us consider the commutative 
limit where $\Theta^{45}\to0$ with the open string coupling $\tGs{M}$ fixed. 
In this limit, the NCYM becomes commutative ${\cal N}=4$ Yang-Mills theory 
in (3+1) dimensions, which shows the well-known electric-magnetic duality. 
On the other hand, we can consider the corresponding limit in the NCOS$_{3+1}$. 
Since, as is seen from \aee, \ogm, and \NCtheta, we have the relation 
$\Theta^{01}=\tGs{M}^2\Theta^{45}$, the commutative limit corresponds to 
the limit $\Theta^{01}\to0$ with the open string coupling $\tGs{E}$ fixed. 
By the relation $\Theta^{01}=2\pi\ae^{(2)}$, we can see that all the massive 
modes of open strings in NCOS theory decouple in this limit. Therefore, 
in this limit, we find that NCOS theory also turns into commutative 
${\cal N}=4$ Yang-Mills theory in (3+1) dimensions. 
We still have the $S$-duality relation \Sduality\ in this limit, 
where the $S$-duality which we have been discussing in this subsection 
turns out to be the ordinary electric-magnetic duality in (3+1)-dimensional 
commutative Yang-Mills theory.

\subsec{OM-theory Compactified on Higher Dimensional Tori}

In the previous subsection, we considered an M5-brane wrapped on a torus. 
In this subsection, we wrap the M5-brane on a circle of radius 
$R_4=f L_4$ in the $x^4$ direction, in addition to the circles in the 
$x^2$ and $x^3$ directions. 

First, we consider the electric picture of this theory, \ie 
we identify the $x^2$ direction as the `11th' direction of M-theory. 
As in the previous subsection, 
the $T$-duality on the circle in the $x^3$ direction maps the D4-brane into 
the D$3$-brane on the circles of the radii $\t R_3$ and $R_4$.
Here the string coupling constant is $\tgs{2}$ and the tension of closed strings 
is $\alpha'_2$.
After taking the $T$-dual on the circle in the $x^4$ direction,
the D$3$-brane becomes a D$2$-brane in type IIA superstring theory compactified 
on the circles of radii $\t R_3$ and $\t R_4={\alpha'_2/R_4}$.
The string coupling constant becomes 
\eqn\tgee{
\gs{4}=\tgs{2} \left(\frac{\t R_4}{R_4}\right)^\hf
={R_2\/fL_3L_4}\sqrt{1\/2R_2\Me^3}, 
}
and the open string coupling is
\eqn\tog{
\Gs{4}^2=\gs{4}f={R_2\/L_3L_4}\sqrt{1\/2R_2\Me^3}. 
}
Therefore, the OM limit (\ie $f\to0$ with $\Me$ fixed) gives 
the NCOS limit in (2+1) dimensions. 

D2-branes wound on the torus in the $x^3$ and $x^4$ directions are the $T$-dual 
of the wound D1-branes on the circle in the previous subsection, 
and the mass of the wound D2-branes equals $1/R_2$. 
In order to decouple the D2-branes with $\ae^{(2)}$ and $\tGs{E}$ fixed, 
we take the limit $R_2\to0$, $\Me\to\infty$, and $L_3L_4\to0$.

In the other picture, we have a D$3$-brane on the circles of the radii $\t R_2$ 
and $R_4$ with $F_{45}$ in \Fm. Using rescaled coordinates 
$x'^i=f x^i$ ($i,j=4,5$), we find that the closed string metric turns into 
$g_{ij}=\delta_{ij}$ and that the gauge field on the D3-brane becomes 
$A'_i=f^{-1} A_i$. Now let us contemplate the $T$-duality on the circle 
in the $x^4$ direction, under which the D3-brane transforms into a D2-brane. 
Since the $T$-duality maps the gauge field $A'_4$ to a scalar field ${X}^{\'4}$ 
on the D2-brane as $A'_4\to (1/2\pi\a_3){X}^{\'4}$, $\Fm$ is 
equivalent by the $T$-duality to 
\eqn\slope{
\frac{f^2}{2\a_3} \partial_5'{X}^{\'4}={L_3\Me^3}. 
}
Using \aem\ and solving \slope, we found that ${X}^{\'4}=f^{-1}{x}^{\'5}=x^5$. 
Thus, the $T$-duality on the circle in the $x^4$ direction maps the D3-brane 
into a D2-brane wound on a line at an angle of $45$ degrees on the 
$x^{\'4}$-$x^5$ plane. 
The radius $\t R'_4$ of the dual circle in the $x^4$ direction is given by 
$1/\t R'_4={R_4}/{\alpha'_3}=2\Me^3L_3L_4$ and the string coupling constant 
is seen to be 
\eqn\mtg{
\tgs{4}=\tgs{3}\left({\frac{\t R'_4}{R_4}}\right)^\hf
=f^\hf{L_3\/R_2L_4}\sqrt{{1\/2L_3\Me^3}}.
}
Then, in the OM limit,  we find that (2+1)-dimensional NCOS theory is 
dual to the theory on the 
D2-brane tilted at an angle of 45 degrees on the $x^{\'4}$-$x^5$ plane
with the Yang-Mills coupling constant
\eqn\YMg{
g_{YM}^2
=\frac{\tgs{4}}{\sqrt{\a_3}}= \frac{L_3}{R_2 L_4}.
}

Note that the D2-brane is spirally wrapping the cylinder in the $x^{\'4}$ and 
$x^5$  directions. The distance between successive turns is finite and nonzero
if we measure it by the canonically normalized scalar field.
Thus, the Higgs VEV of the effective theory on the D2-brane
becomes finite. Note also that, by lifting the D2-brane to M-theory, 
we can see that this theory is 
equivalent to the description of the ${\rm NCOS}_{2+1}$ 
made by using an M2-brane in \GoMiSeSt, except for the fact that 
this theory is compactified on a 3-torus. 
In particular, the eleven-dimensional scale and the radius of the circle
along the $x^4$ direction are coincident.
The difference made by the compactifications may disappear in the NCOS limit
since the radii of the circles are very large in the NCOS theory.

Here, we will discuss D0-branes in $(2p+1)$-dimensional NCOS theory. 
D0-branes have finite mass 
\eqn\Dmass{
M_{\rm D0}=\frac{1}{\g \sqrt{\alpha'} }= 
\frac{1}{\G^2 \sqrt{\alpha'_{\rm eff}} },
}
where $\G$ is the open string coupling constant in NCOS theory. 
From \Dmass, we can see that D0-branes do not decouple 
from the spectrum of NCOS theory until we take $\G$ or $\ae$ to zero.
However, we can find that these D0-branes are bounded in the $D(2p)$-branes, 
since, if the D0-branes were outside the $D(2p)$-branes, the 
mass of the tachyon field in the worldvolume theory of the D0-brane
would be proportional to ${\alpha'}^{-\frac{1}{2}}$, as shown in \SW.
In the D4-branes ($p=2$) case, the bounded D0-brane corresponds to an instanton 
in the low-energy Yang-Mills theory. 
Thus, the D0-branes may play a role as a nonperturbative object 
in $(4+1)$-dimensional NCOS theory.
In the D2-branes case, a uniform electric flux on the worldvolume corresponds 
to the bounded D0-branes, which preserve the same number of supersymmetries. 
Since the coordinates $x^0$ and $x^1$ are not compactified, any finite electric 
flux cannot produce a finite number of bounded D0-branes. 

Now, let us move to OM-theory compactified on $(q+1)$-dimensional 
spatial torus $(0 \leq q \leq 3)$,
where the coordinates $x^2$ and $x^j \, (3\leq j\leq q+2)$ are periodically 
identified and their radii are $R_2$ and $R_j=f L_j$, respectively.
By identifying $R_2$ as $R$ and applying the $T$-duality to the $q$ circles,
we find that 
the OM limit becomes the ${\rm NCOS}_{(4-q)+1}$ limit and that 
the closed string coupling constant is given by $\g=\G^2/f$, where 
\eqn\og{
{\G^2}
=\frac{R_2 }{\prod_{j=3}^{q+2} L_j}{{\alpha'}_{\rm eff}^{\frac{q-1}{2} } } 
}
is the open string coupling constant. 

In the ${\rm NCOS}_{(1+1)}$ case, 
the open string coupling is
$\G^2
={\ae R_2}/(L_3 L_4 L_5)$.
In order to decouple D3-branes wound on $\T^3$ in the $x^3$, $x^4$ and $x^5$ 
directions with mass $M_{\rm D3}=1/ R_2$ and D1-branes wound on the circle in 
the $k$ direction ($k=3,4,5$) with mass 
$M_{\rm D1}={\epsilon^{ijk} L_i L_j}/(R_2 \ae)$, we need to take 
the limit $R_2, L_3, L_4, L_5 \rightarrow 0$ with ${R_2}/(L_3L_4L_5)$ fixed.
In a dual picture in which $R$ is identified as  $R_3$,
the magnetic flux $F_{45}$ must be quantized as
\eqn\magflux{
2 \pi n=\int_{0}^{2 \pi L_4} \int_{0}^{2 \pi L_5} 
dx^4 dx^5 F_{45}=4 \pi L_3 L_4 L_5 M_{\rm eff}=2 \pi {\G^{-2}}.
}
However, under $T$-duality on the circle in the $x^4$ or $x^5$ direction, 
the tilted D2-brane transforms into a D3-brane, not a D-string. 
Although we could map the D2-brane to a D-string under $T$-duality, 
instead of doing that, in the next section 
we will give a simple description of the $S$-dual theory of 
NCOS$_{1+1}$ by using an M2-brane.

\newsec{(1+1)-dimensional NCOS theory and M2-branes}

In this subsection, 
we consider an M$2$-brane in M-theory compactified on a rectangular 
torus, which is a product of circles with circumferences $2\pi R_{11}$ 
and $2\pi fL_{2}$, with a metric $g_{\mu\nu}=\eta_{\mu\nu}$ 
($\mu, \nu=0,1,11$) and $g_{22}=f^2$:  
\eqn\del{
f^2=\frac{2\Me^3}{\M^3}, 
}
and so the coordinate $x^2$ is identified under 
the translation by $2\pi L_2$ as $x^2\sim x^2+2\pi L_2$. 
Let us suppose that the M2-brane worldvolume is of the form 
$\R^2\times\S^1$. For the $\R^2$, we take coordinates $x^0$ and $x^1$. 
The circle $\S^1$ is tilted at an angle of 45 degrees on the 
$x^2$-$x^{11}$ plane, where we assume that $L_{2}= n R_{11}$ and that 
the M2-brane is wound once on the circle in the $x^{2}$ direction. 

According to the dictionary between M-theory and type IIA theory, 
the scalar field $X^{11}$ $(=2\pi R_{11}\phi)$ on the M2-brane worldvolume 
is the electric-magnetic dual of the gauge field in the (2+1)-dimensional 
DBI action of a D2-brane in type IIA theory. The scalar field $\phi$ has 
the usual normalized kinetic term and is mapped to the gauge field $A$ as 
\eqn\elemag{
\pder{x^{\'2}}\phi={1\/2\pi\g\a^\hf}{2\pi\a F_{01}\/\sqrt{1-(2\pi\a F_{01})^2}}, 
}
where we use the coordinate $x^{\'2}$, which is given by rescaling $x^2$ as 
$x^{\'2}=f x^2$ and so is identified under the translation by 
$2\pi R_2=2\pi fL_2$. Since the circle on which the M2-brane is wound tilts at 
an angle of 45 degrees on the $x^2$-$x^{11}$ plane, \ie $\S^1$ on the line 
$x^2=x^{11}$, we can see from \elemag\ that there is nonzero background 
electric flux 
\eqn\eleF{
2\pi\a F_{01}\sim1-\hf f^2 \qquad (f\ll1)
}
on the D2-brane worldvolume. 
Thus, this system of the M2-brane is equivalent to a D2-brane with 
$F_{01}\neq0$ in type IIA theory with ${1}/{\alpha'}=R_{11}\M^3$ and 
$\g^2=(R_{11}\M)^3$.
Thus, the OM limit $\M\to\infty$
is the (2+1)-dimensional NCOS limit with the open string coupling constant 
$\G^2=\sqrt{2}(R_{11}\Me)^{3\/2}$ \GoMiSeSt.

Applying $T$-duality to the circle in the $x^2$ direction, we map 
the D2-brane in the previous paragraph to a D1-brane with the electric 
flux in \eleF\ in type IIB theory compactified on a circle of radius 
$\tilde{R_2}={\a}/{R_2}={f}/({2nR_{11}^2\Me^3})$, 
with the string coupling constant $\tg=\g\sqrt{\a}/{R_{2}}=1/(nf)$ 
and the effective string tension $1/\ae=f^2/\a=2R_{11}\Me^3$. 
As for the open string moduli, we have the noncommutative parameter 
$\Theta^{01}=2\pi\ae$ and the open string metric $G_{\mu\nu}=f^2\eta_{\mu\nu}$ 
($\mu, \nu=0, 1$), from which we can see that the open string coupling constant 
is $\tG^2=f\tg=1/n$. 

In the OM limit $\M\to\infty$ with $\Me$ fixed,
this system goes to (1+1)-dimensional NCOS theory 
with the open string coupling constant $\tG^2={1}/{n}$ and the effective string 
tension $1/\ae=2R_{11}\Me^3$.
The mass of D-strings wound on the circle in the $x^2$ direction is 
$M_{\rm D1}={\tilde{R_2}}/({\tilde{g}_{s} \alpha'})={1}/{R_{11}}$. 
In order to decouple these D-strings with $\ae$ fixed, we can take the limit 
$R_{11}\to0$ and $\Me\to\infty$.

On the other hand, thinking of the $x^2$ direction as the `11th' direction of 
M-theory, we can view this system as $n$ D2-branes with $F_{01} \neq 0$ 
in type IIA theory with the string tension 
${1}/{\alpha'_2}=R_{2}\M^3=2nR_{11}\Me^3/f$ 
and the string coupling constant 
$\gs{2}=({R}_{2}\M)^{3\/2}=\sqrt{2f}(nR_{11}\Me)^\frac{3}{2}$.
For the $n=1$ case, the line $x^2=x^{11}$ on which the M2-brane is wound 
corresponds to nonzero electric flux on the worldvolume of the D2-brane. 
It is convenient to normalize the scalar field $X^2$ on the M2-brane so that 
the metric which appears in the kinetic term of 
the scalar fields becomes $\eta_{AB}$. Therefore, we have the relation 
$X^2=fx^{11}$ (\ie $\d_{11}X^2=f$). As in the previous picture, 
by making use of 
\eqn\telemag{
{1\/2\pi R_2}\pder{x^{11}}X^2={1\/2\pi\gs{2}{\a_2}^\hf}
{2\pi\a_2 F_{01}\/\sqrt{1-(2\pi\a_2 F_{01})^2}}, 
}
we find that the electric flux on the D2-brane worldvolume is 
\eqn\FE{
2 \pi \alpha'_2 F_{01}\sim f \qquad (f\ll1).
}
For general $n$, the flux may be obtained as $2\pi\a_2F_{01}\sim f\one_{n}$.

Taking the $T$-dual on the circle in the $x^{11}$ direction, 
we obtain $n$ D1-branes with the electric flux in \FE\ in type IIB 
superstring theory compactified on the circle of radius 
$\tilde{R}_{11}={\alpha'_2}/{R_{11}}=f/(2nR_{11}^2\Me^3)=\tilde{R}_{2}$
and the string coupling constant 
$\tgs{2}=\gs{2}\sqrt{\a_2}/R_{11}=nf=1/\tg$.
In the OM limit, we see that $\a_2\to0,\,\tgs{2}\to0$, and that 
$2\pi\a_2F_{01}\to0$. 
Therefore, the OM limit can be identified as the YM limit, where 
this theory becomes (1+1)-dimensional $U(n)$ Yang-Mills theory with 
a single unit of electric flux and the coupling constant 
\eqn\YMG{
\frac{1}{2\pi g_{\rm YM}^2}= \frac{\a_2}{\tgs{2}}=\frac{\ae}{n^2}.
}
Indeed, this is the $S$-dualized theory of (1+1)-dimensional NCOS theory 
\GoMiSeSt. In this case, to decouple fundamental strings wound on the circle, 
which are the $S$-dual of the wound D-strings and have mass $1/R_{11}$, 
we have to take the same limit in the dual picture.

\newsec{Summary and Discussion}

In this paper, we have studied the compactification of OM-theory on tori, 
with the effective Planck scale $\Me$. As shown in \GoMiSeSt, 
depending on which circle is identified as that in the `11th' direction 
in M-theory -- whether the direction tangent to the circle is aligned with 
that of nonzero $H_{0ij}$ ($i$ denotes the spatial direction) or not -- 
OM-theory compactified on the circle 
becomes NCOS theory or NCYM theory in (4+1) dimensions. We have shown that 
the compactification of OM-theory on higher dimensional tori, which 
corresponds to the $T$-dual of the NCOS$_{4+1}$ on tori, gives 
lower dimensional NCOS theory where the effective string tension 
$1/\ae=2R\Me^3$ and the noncommutative parameter $\Theta^{01}=2\pi\ae$. 

From section 2, even after taking the NCOS limit, we still have `$T$-duality' 
between NCOS theories. 
NCOS$_{(p+1)+1}$ theory from a D$(p+1)$-brane wound on a circle is mapped 
to NCOS$_{p+1}$ theory from an unwound D$p$-brane under $T$-duality on the 
circle. Let us suppose that the circle in NCOS$_{(p+1)+1}$ theory has 
a coordinate $s$, which is identified under the translation $s\to s+2\pi L$. 
Before taking the NCOS limit $f\to0$, we assume a metric $g_{ss}=f^2$ on the 
circle. The $T$-dual of this circle is a circle with a coordinate $\tilde{s}$ 
identified under $\tilde{s}\to\tilde{s}+2\pi\t{L}$ and with the metric 
$g_{ss}=f^2$. Then, the open string coupling constant $\G^2$ in NCOS$_{(p+1)+1}$ 
is mapped under the $T$-duality to $\tG^2$ in NCOS$_{p+1}$ as 
\eqn\Tdual{
{\sqrt{L}\/\G^2}={\sqrt{\tilde L}\/\tG^2}, \qquad 
{\rm with} \qquad \tilde L={\ae\/L}.
}
By making use of \Tdual, we can reproduce all the open string coupling constants 
$\G^2$ in the NCOS theories which we studied in this paper. 

In sections 2 and 3, we have demonstrated that we can understand, 
from the OM-theoretical point of view, what the $S$-duals of the NCOS theories 
are, although the $S$-dualized theories have already been identified in 
\GoMiSeSt. In particular, in (3+1) dimensions, we have seen that 
OM-theory compactified on a torus gives a simple heuristic derivation of 
the $S$-duality between NCOS$_{3+1}$ and NCYM$_{3+1}$. When the torus is the 
form of a product of circles of radii $R$ and $fL$, the open string coupling 
$\tGs{E}$ in NCOS$_{3+1}$ is given by $\tGs{E}^2=R/L$ and $\tGs{M}$ in 
NCYM$_{3+1}$ by $\tGs{M}^2=L/R$. Thus, the $S$-duality can easily be understood 
as the exchange $L\leftrightarrow R$. Also, it is interesting to notice that 
the electric flux $F_{01}$ in the NCOS theory and the magnetic flux $F_{23}$ 
in the NCYM theory are unified into the self-dual three-form field $H$ in 
the M5-brane worldvolume theory.

\vskip 0.5in
\noindent
{\sl Note added}: while writing this paper, we received the paper \BBSS, 
which has some overlap with ours in the discussion of $S$-duality between 
NCOS and NCYM in (3+1) dimensions. Also, upon completion of this manuscript, 
we received the related papers \refs{\RS,\LRS}.

\vskip 0.5in
\centerline{{\bf Acknowledgements}}
We would like to thank Tsuguhiko Asakawa and Isao Kishimoto for discussion.
T.$\,$K. was supported in part by a Grant-in-Aid (\#11740143)
and in part by a Grant-in-Aid for Scientific Research 
in a Priority Area: ``Supersymmetry and Unified Theory of Elementary 
Particles''(\#707), from the Ministry of Education, Science, Sports and 
Culture.
S.$\,$T. was supported in part by JSPS Research Fellowships for Young
Scientists.

\listrefs

\end